# Design of mechanically robust metasurface lenses for RGB colors

Jun Yuan, Ge Yin, Wei Jiang, Wenhui Wu, and Yungui Ma*

*State Key Lab of Modern Optical Instrumentation, Centre for Optical and Electromagnetic Research, College of Optical Science and Engineering, Zhejiang University, Hangzhou 310058, China*

*Corresponding author's email: yungui@zju.edu.cn

## ABSTRACT

In recent years, dielectric rod based metasurface lenses have been particularly investigated for their potential applications in replacing the traditional bulky lens with high efficiency. However, the isolated granular structure may lead to robustness and chromatic aberration concerns. In this work, a low refractive-index embedding medium was applied to solve the structural stability problem that also enables the device to be transferable to desired position or surface. Based on this, a compound metasurface lens composed of randomly interleaved frequency-selective zone sectors is proposed to broaden the bandwidth of the two-dimensional device. The validities of the proposed method and the application potentials for the multifunctional lens that can manipulate RGB three colors have been numerically examined and verified. The current results are of essence in guiding the future design of metasurface lenses for real practice.

Metasurface as initially introduced to generalize the Snell's law, which is actually a special case of the grating equation, has attracted great research interests in the field of metamaterials. It enables the manipulation of electromagnetic wave propagation with a gradient geometrical phase profile[1-6]. The two-dimensional artificial structure composed of arrays of subwavelength scatters could produce anomalous reflection and refraction phenomena[7, 8], which is practically more promising in realization compared with the three-dimensional ones. The optical functionalities of metasurface have been deeply explored and various applications have been proposed which include optical lenses[9, 10], holograms[11-15], wave beam transformers[16], or serve as unique platforms to generate various optical polarization/momentum states[17-20], computation devices[21, 22], digital electromagnetic (*EM*) surfaces[23-25], *etc*. Among them, the metasurface lens is mostly investigated with the expectation to replace the traditional bulky lens by a piece of structured surface layer, which may hopefully bring about significant advancements for modern optical systems[26-32]. Compared with the existing thin film based Fresnel's lens that is mainly to yield a required phase profile, the metasurface lens can deal with both transmission magnitude and phase and even polarization in subwavelength scale resolutions. The increased freedoms will broaden the functions of the lens and more importantly improve the focusing/imaging capability, for example by suppressing undesired diffractions or reflections. The fabrication process potentially enabled by a single step lithography is a technical advantage that is vital for the final production[10, 31]. To improve the efficiency of metasurface lens, dielectric materials with high refractive index have been developed to suppress the loss, such as Si used for infrared and visible light[10, 30, 33, 34], $TiO_2$ for visible light[31, 32, 35] and $SiN_x$[36, 37]. A bright future for metasurface lens could be envisioned by putting all these improvements into perspective.

However, for applications, there are still some practical issues needing to be managed, among which the device robustness and bandwidth are mostly concerned. The isolated dielectric rods as experimentally used in the samples will have the stability problems against mechanical, thermal or chemical agitation, and the diffractive nature of the spatially distributed elements will also lead to large inherit chromatic aberrations[32, 38]. These issues are truly vital when considering an imaging lens especially for visible light. Multi-wavelength functional metasurface lenses have been explored by different groups[9, 34, 38-46]. They utilized either polarizations of two-wavelength light[40, 43], or optimization algorithm to broaden the working wavelength range[9, 39]. For the former case, the two wavelengths are usually purposely selected to have large difference so that their unit structures can have very low electromagnetic interference[38, 44]. Ostensibly, this approach will be challenged to realize multiple colors for visible light, in particular regarding the issues such as cross-talk among different color channels[42], or multiple spatial focusing for single color[34, 45]. The reflection type metasurface proposed in Ref. 41 may hopefully reduce the interference among different channels in visible light, but the practical application is constrained by low reflectivity. In this letter, aimed to solve these problems, we propose a general approach to design a solid metasurface lens working for multiple colors, which in principle could be extended for any frequency regime. Low index dielectric claddings for metasurface devices have been explored before to acquire scalability and flexibility[29, 30, 47-51]. But the effect of their structural features to the device performance has been rarely discussed. In this paper we studied the influences of thickness of the embedding medium and the spatial positions of the embedded rods with respect to the transmission coefficient. We will show that a stable and transferable metasurface lens can be configured by embedding the dielectric resonance rods in a low refractive index medium (Teflon AF 2400) without scarifying the optical modulation freedoms. Based on the embedding structure, we design an RGB-color functional lens (*i.e.*, hopefully working for white light sources) by developing a multifunctional metasurface[20, 52, 53]. Different color sectors are randomly interleaved to control the undesired diffraction. The validities of the approach and the devices proposed here have been numerically verified by full-wave simulations that exhibit desired focusing effects at different wavelengths. The imaging performance for a color picture is also discussed based on the predefined RGB channels. These results are instrumental by providing a promising blueprint for the design of a real practicable metasurface lens for visible light.

**Metasurface design**

Figure 1a gives a schematic of the metasurface lens proposed here, which can primarily work for three different colors. In this paper, we select red, green and blue colors at the wavelength of 700, 550 and 420 nm, respectively. The main part of our optical lens is composed of $TiO_2$ rod arrays. By this lens, we expect that the normal incident plane-waves at these three wavelengths can be focused at the same locations, *i.e.*, having the same focal length. This function is hardly available by the general metasurface lens design that usually considers a single wavelength[31, 32]. As will be addressed in more details later, an interleaved multi-wavelength phase profile is introduced here for this purpose. First with respect to the robustness issue, a transparent matrix with low refractive index is directly considered to embed the dielectric resonators and isolate them from the surrounding environment, and the lattice type is square, as schematically shown in Figs. 1b-d. Here the polymer Teflon AF 2400 with index $n = 1.29$ in the visible light has been introduced to sever this purpose[54, 55]. The embedding approach will also allow the metasurface lens to be possibility separated from the processing hard substrate to be a real single layer or transferred to other desired positions, which definitely will be very essential for many practical applications.

Recently, metasurface lenses composed of $TiO_2$ rods arrays in the air background have been demonstrated for visible light[56]. In the supplementary information, we give a thorough investigation how the transmission coefficient (magnitude and phase) of the $TiO_2$ rod periodical array with the Teflon matrix changes with the structural parameters (rod diameter *d* and period *p*) at the desired RGB colors. The permittivity spectra of $TiO_2$ are obtained from Ref. 57. We find that by controlling these structural parameters, it is still possible to achieve a large average transmissivity (> 90%) and simultaneously cover a $2\pi$-phase variation, as example shown in Fig. 2a ($p = 200$ nm), b ($p = 350$ nm) and c ($p = 400$ nm) for the three wavelengths, respectively. In these figures, as indicated by the solid lines of different colors, we see that increasing the thickness of the embedding medium to be larger than the height of the $TiO_2$ rod will modulate the transmission spectra but will not change the overall transmissivity values that are mostly above 80%. The transmission phase defined with the reference planes coincident with the two end surfaces of the dielectric rods is also not disturbed with the increase of the embedding medium thickness. It shows that the thickness of the embedding medium has little impact on the transmissivity and phase. These results are practically important by affording the experimental freedom in controlling the thickness of the embedding medium. For each wavelength, there are some different heights of $TiO_2$ could be chosen. Usually larger working wavelength requires the lens with higher rod arrays. Here, we chose 600 nm because this sized $TiO_2$ rod could realize the desired $2\pi$ phase variation in the visible light region we are interested by tuning the radius. Rod height much larger than 600 nm will cause undesired dielectric resonances at shorter wavelengths such as 420 nm. The diameters for the dielectric rods suggested here are larger than 50 nm, which is within the fabrication capability of modern electron beam lithography[31, 42]. Different from the elliptical rods or nanofins used before[27, 58], the circular rods used here is polarization-independent and will be more favorable for imaging applications [10, 59, 60].

In the following, we proceed to design the metasurface lens by the transmission and structural data given in Figs. 2a-c. Our goal is to build a compound metasurface lens functional for the RGB colors, which has diameter $D = 1.12$ mm and focal length $f = 0.42$ mm, corresponding to a numerical value NA = 0.8. With these predefined parameters, we can decide the arrangement of the dielectric rod array in the metasurface and the transmission phase profile using the simulated data. Based on these information, the transmitted field distribution is analytically computed in terms of the Fresnel diffraction equations[61]. More detailed descriptions about the calculation are given in the supplementary information where we also give the focusing spot cross-sections and their intensity line profiles of the metasurface lens individually designed for different colors. Light focusing performance with spot sizes of diffraction limit is obtained for these lenses. But in frequency, neither of them can cover the whole visible region without causing serious chromatic aberrations. Inspired by the recent work on multifunctional metasurfaces[20], we may combine the three different lens into a single design thus realizing a multi-wavelength functional metasurface lens. For imaging application, it is essential to keep an aperture size as large as possible. For this reason, the interleaved distribution scheme for the lens sectors with each of them controlling different colors is adopted here and carried out by two types of arrangement methods. As shown in the top row of Fig. 3a, the first one is the periodical arrangement for the three RGB color sectors, which is most straightforward and convenient. The total sector number $N$ is a variable that could be tuned to control the focusing quality and fabrication efficiency. Three different band-pass filters that are only transparent for desired wavelengths will be assumed covering the sectors, which practically could be realized by depositing colorful photoresists as used in modern color sensors[62]. In the bottom row of Fig. 3a, we plot the field profile at the focal plane at the wavelength of 700 nm. Well-defined focusing spot is obtained but accompanied by high diffraction order spots coming from the periodical color zones. Increasing the lens sector density can reciprocally separate the different order spots to have larger distances, as clearly shown from the intensity line curves plotted in Fig. 3b.

As the second method, a random deployment for the lens sectors that have the same size and same numbers for each color has been applied to minimize the undesired diffraction. As shown in Fig. 3c and d, a single focusing spot is attained and the quality is improved with the increase of the sector density. When $N = 2500$ (the side length of the square color sector equal to 22.4 μm), the intensity profile of the focus spot already has almost no difference with that of a single channel case (the whole lens is solely designed for 700 nm incident light). We calculated the transmissivity and focusing efficiency of the RGB functional metasurface lens with the sector number $N$. The results are plotted in Fig. S4 in the supplementary information. The transmissivity values, the intensity ratio of the outgoing and incident waves of the lens, for different color channels fluctuate around 33% for both periodic and random arrangement lenses due to the small number difference for each color, which is more obvious at smaller $N$. The rest energy is assumed to be absorbed by the filters. The focusing efficiency of the lens was calculated by the ratio of the energy flow integrated over the focal spot area to that passing through the focal plane of the lens. For the lens of periodical sectors, the focusing efficiency for different color channels are about 28% at $N \geq 100$, which could be calculated according to the grating equation and the diffraction theory as shown in the supplementary information. For the lens made of random color sectors, which mimics aperiodic gratings, light waves that form higher order focusing spots in Fig. 3a will now be randomly distributed as background noise in the focal plane with a spatial spreading scope solely determined by the unit sector size. This background diffraction will become much broader and weaker as $N$ increases and their influence to the desired (zeroth-order diffraction) focusing spot is negligibly small at $N = 2500$. More discussions are given in the supplementary information. Obviously, the second random phase profile is effectively more favorable although it may raise the fabrication complexity. Figs. 4a-c show the focusing effect for normal incident plane wave at different colors by the logarithmic electric field intensity profile near the focal point at $N = 2500$. Similar results with the same focal length for the three colors are obtained. The half width of the Airy function form focus spot (Figs. 4d-f) is 245, 345 and 417 nm for the operating wavelength of 420, 550 and 700 nm, respectively. Calculated from $0.514\lambda/\text{NA}$, the theoretical diffraction limited FWHM should be 270 nm, 353 nm, and 450 nm for the three wavelengths considered here. As described in the supplementary information, the grid spacing (mesh) applied in our calculation is 25 nm. This size limit may lead to a largest uncertainty error about the estimated FWHM at about 50 nm. We will not expect the metasurface lens can break the diffraction limit. Actually, we further calculate the radius of the first dark ring of the Airy disk, which are 230 nm, 336 nm and 375 nm, larger than the diffraction limit calculated from $1.22f\lambda/d$ (193 nm, 252 nm and 321 nm) for the three RGB colors.

## Results

**Focal profile of metasurface.** In the above we have analytically proposed a metasurface lens that could work for three different RGB colors. Although the implementation will be a challenging task, the practical potential could be first examined by numerical simulation. In Figs. 5a-c, we give the full-wave simulated electric-field pattern of the metasurface lens (thickness $h = 0.8$ μm) designed for the normal incident plane wave at wavelength $\lambda_0 = 420$, 550 and 700 nm, respectively. To prove the potential and also for simplicity, the lens structures are set to be periodically invariant along the $y$-axis and have an aperture $D_0 = 25.2$ μm along the $x$-axis. The structural and material parameters for the lens elements are picked from the databank plotted in Figs. 2a-c at a predefined focal length $f$ = 9.4 μm (NA = 0.8). Well-shaped focusing behaviors are evidently shown by these pictures. One-dimensional cuts of theses focus patterns are plotted in the supplementary information. The energy transmission ratio calculated along the focus line is larger than 79% as realized by the precise design and control of each element in the metasurface. In the multi-wavelength RGB lens proposed above, there will be numerous distributed lens sectors with similar light manipulation capabilities as shown in Figs. 5a-c. To mimic this effect, we partially cover the metasurface lens by pieces of perfect electric conductor (PEC) films (width = $2D_0/9$) that play the role of a band-pass filter by leaving three uncovered working regions that have a same aperture size = $D_0/9$. As shown in Figs. 5d-f, the directional focusing effect is basically maintained for the light passing through the opened metasurface regions. The broadening of the focus spot in particular at larger wavelengths is caused by the irregular diffraction occurring for the reduced apertures, which is only 2.8 μm in our simulation. Moreover, insufficient number of color sectors is also an important reason. This practically could be controlled by increasing the area and number of color sectors, as for example that the minimum zone sectors proposed above at $N$ = 2500 is 22.4 μm roughly equal to the lens aperture modeled in Fig. 5a-c.

We have introduced low-index embedding medium to improve the physical robustness of the metasurface sample. This operation has limited effect on the resonance characters of the $TiO_2$ rods and the dispersion of their arrays. For traditional lens systems, complex combination of main and accessory optical components has been generally built to correct various aberrations. To exhibit the advantages of the metasurface design, we employ a multifunctional conception to control the chromatic aberration but at the price of reducing the received energy. This compound design is very important, in particular when imaging a color scene with a shallow depth of field. To improve both energy utilization efficiency and image quality, the band-pass filter as proposed here is expected to have a working width around 140 nm and three of them could cover the whole visible light spectrum (380-780 nm) as many as possible. Technically, it could be realized by colorful photoresists as widely used in modern charge-coupled device (CCD) color sensors. Figure 6 shows the calculated focal length distributions in the three RGB color windows by considering the actual dispersion of each element in the different metasurfaces. It is seen that by our design, the maximum variation for the focal length is narrowed down to ± 50 μm, which is about three times smaller than that of a single channel lens.

**Imaging profile of metasurface.** It will be interesting to see the quality of our compound metasurface lens in imaging a real RGB color picture, which is numerically examined in the last part. Figure 7a shows a flower that could be encoded as the mixture of red (Fig. 7b), green (Fig. 7c) and blue (Fig. 7d) three pictures. We place this flower picture at the two focal-length distances from the metasurface lens (consisting of randomly distributed color sectors with the total number $N = 2500$, aperture $D = 1.12$ mm and focal length $f = 0.42$ mm). At the equal distance of other side, an inverse color image (Fig. 7e) could be observed by a screen, while the picture imaged by each channel is draw in Figs. 7f-h for different colors. The computation process is described in the method section. We see that the image information is exactly captured and reconstructed by the compound metasurface lens except for the great reduction in the image intensity mostly due to the application of band-pass filters, which in principle could be improved by adopting a lens with larger aperture. The numerical experiment unambiguously indicates the practical potential of this RGB metasurface lens in dealing with the incidence of visible light with arbitrary wavelength.

## Discussion

Metasurface lenses made of low-loss dielectric resonators have been expected to probably generate the first real optical applications for metamaterials, which will be more practically solid by adopting the composite embedding structure as proposed in this work. This is particularly necessary when considering the situation that the dielectric rods of high aspect ratios are usually fabricated on alien substrates. An additional spin-coating process for the embedding medium may be enough as we show the effect of the embedding layer to the optical properties is very small. The compound lens structure integrated with multiple frequency-selective working regions provides an ingenious solution to tackle the chromatic aberration problem and the randomly interleaved sectors are important to maintain the aperture size without causing additional diffraction. Compared with the previous reports, usage of wavelength filters for different color sectors could avoid undesired focal spots and spectral interference among different color channels[34, 42]. And also allow the equal distribution of incident energy to each channel. Besides, the working wavelength could be more freely selected when a top filter is used, which is practically very favorite. The validities of the proposed method and the metasurface lens are numerically examined and verified. These results are instrumental in providing a clear guidance for the further development of metasurface-based lens applications.

**Methods**

**Numerical simulation.** (i) For the transmission spectra given in Figs. 2a-c, finite-element frequency-domain method is carried out to simulate the transmission spectra using COMSOL Multiphysics. In the simulation, the permittivity of $TiO_2$ at 420, 550, 700 nm is 10.8, 8.7 and 8[57], respectively and the refractive index of Teflon AF 2400 is 1.29 in the range of visible light[54, 55]. Periodic boundary condition is set around the structure. We change the thickness of surrounding media (from 0 nm to infinity) to study its influence on the transmissivity and transmission phase of structure; (ii) For the field patterns given in Figs. 5a-f, the finite-integration frequency-domain technique is carried out to simulate the field distribution using CST Microwave Studio. In the simulation, the diameter of the lens is 25.2 μm, in order to make sure the integer multiple of periodic structures at all three working wavelengths. The focal length is 9.4 μm for a numerical aperture of 0.8. According to the phase distribution calculated on the metasurface, the radius value is selected from the phase-radius relations computed in (i).

**Analytical calculation**. (i) For a single-channel metasurface lens, the lens is illuminated by the normal plane wave. We could get the intensity of light distribution on the focal plane through the convolution between the lens of matrix and the Fresnel diffraction factor whose propagation distance is equal to the focal length; (ii) For the RGB multi-wavelength lens, different band-pass filters are placed on the metasurface lens to define the incident light wavelength. The transmissivity of the filter is set to be unit (zero) when the incident light is in (out) the working wavelength range. Different colored light propagates through the lens through different filter regions; (iii) For the inspection of imaging quality of the RGB lens, a colorful picture is placed at the two times focal-length distances from the metasurface lens as an object. Three monochromatic RGB component pictures are separately calculated at the imaging side. The final image is the superposition of these three monochromatic images. The calculation process is introduced in the supplementary information.

**Acknowledgments**

The authors are grateful to the partial supports from NSFC 61271085 and NSFC of Zhejiang Province (LR15F050001).

**Author Contributions**

J.Y. and Y.M. conceived the design; J.Y. did the numerical and analytical analysis; G.Y. assisted in simulations; All the authors discussed the results and commented on the manuscript; Y.M. supervised the research.

**Additional Information**

**Supplementary information:** accompanies this paper at http://www.nature.com/srep

**Competing financial interests:** The authors declare no competing financial interests.

**Figure 1 by Yuan *et al.***

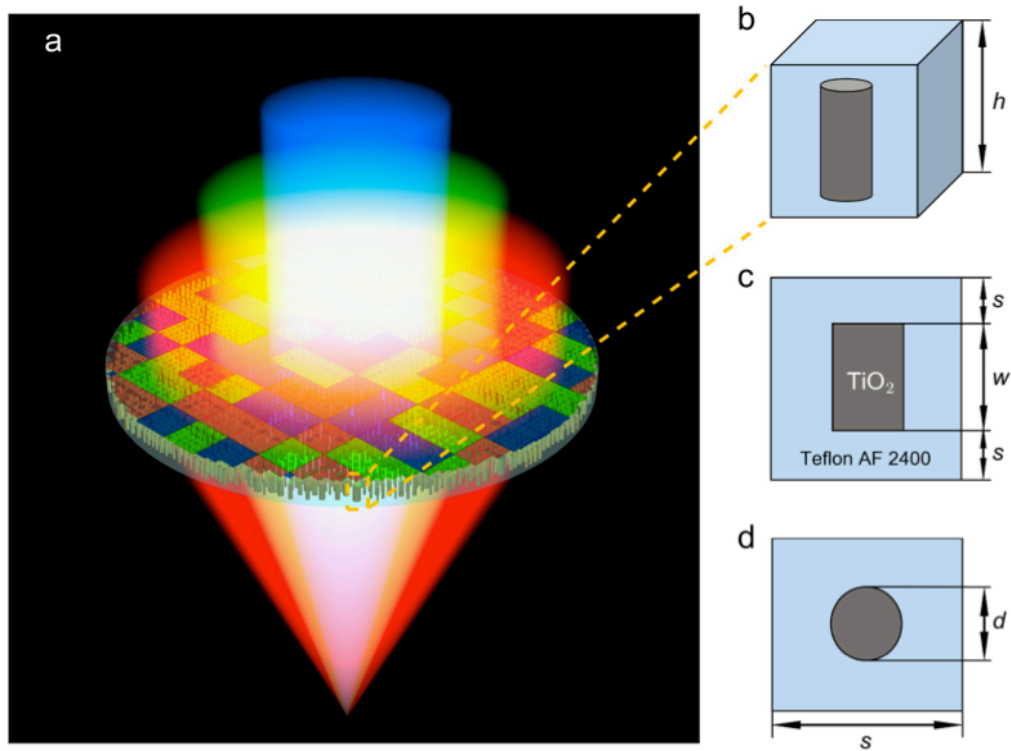

**Figure 1. Schematic of the RGB colors functional metasurface lens.** (a) The lens structure and the property with equal focal lengths for selected red, green and blue colors. The lens is composed of TiO$_2$ rods fillers embedded in Teflon AF 2400 matrix that are covered by different band-pass filters as indicated by colorful rectangles. (b)-(d) Different views of the metasurface's element together with the structural parameters. In the later simulation, we tune the period *p* of the unit cell and the diameter *d* of the rod to acquire the desired transmission coefficient.

**Figure 2 by Yuan *et al.***

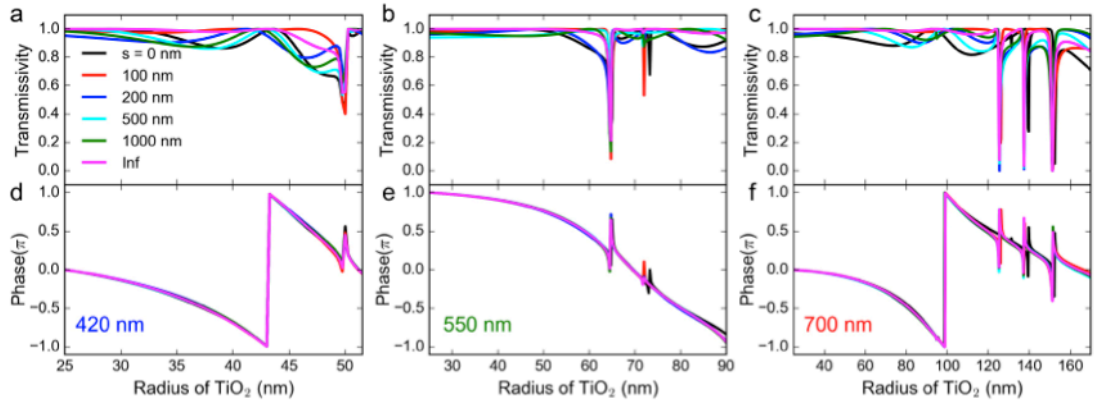

**Figure 2. Transmission coefficient of the periodical TiO₂ rod array changed with the radius of the rod and the thickness of the embedding medium.** (a)-(c) Transmissivity (top) and phase (bottom) curves calculated for the operation wavelength of 420, 550 and 700 at $p$ = 200, 350 and 400 nm, respectively. The height of the TiO₂ rod is fixed at 600 nm. The parameter 's' as defined in Fig. 1c indicates the surface distance between the embedded rods and embedding medium. In the phase calculation, the reference planes are set to be coincident with the two ends of the rod arrays. The sharp transmissivity dips are caused by the dielectric resonance that should be technically avoided in the lens design.

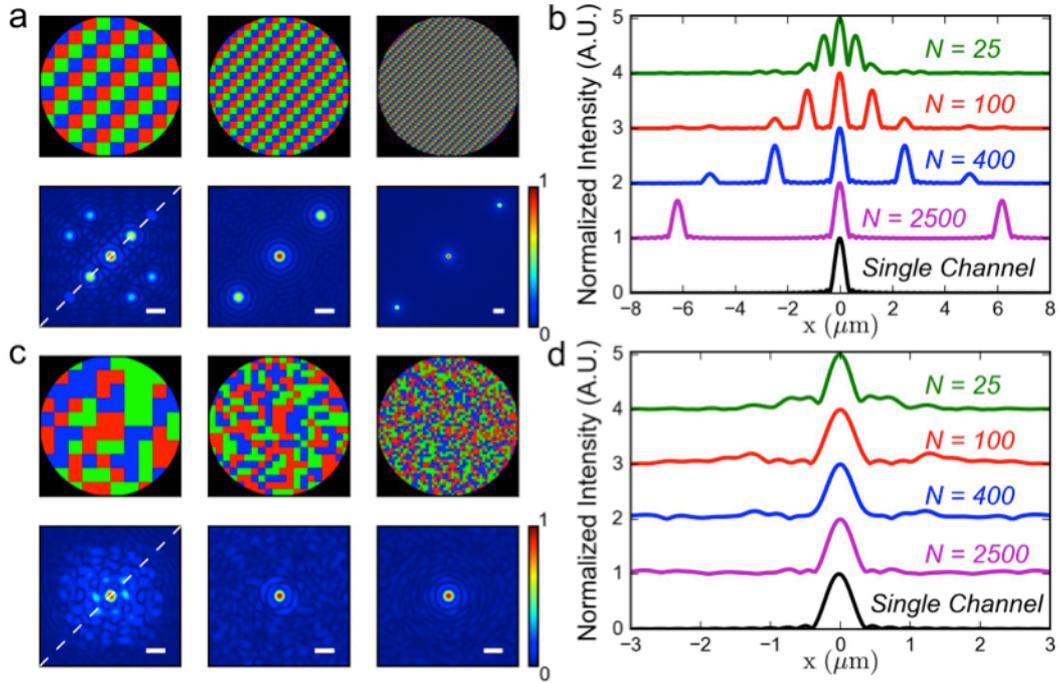

**Figure 3. Design of the RGB colors functional metasurface lens. Here the lens's aperture, focal length and NA are 1.12 mm, 0.42 mm and 0.8, respectively.** (a) and (b) The RGB color sectors are periodically deployed. The top row from the left to right side in (a) shows the distribution of color sectors with the total number $N = 100$, 400, 2500 and the corresponding field intensity in the focal plane is drawn in the bottom row. The curves in (b) plot the intensity profile along the dashed line as indicated in (a) for the lens of different sector numbers. 'Single channel' means the case with the lens solely designed for a single frequency. The white lines in the field intensity pictures indicate the length of the operation wavelength. (c) and (d) The RGB color sectors are randomly deployed. The descriptions for the pictures in (c) and (d) are the same as those appearing in (a) and (b).

**Figure 4 by Yuan *et al.***

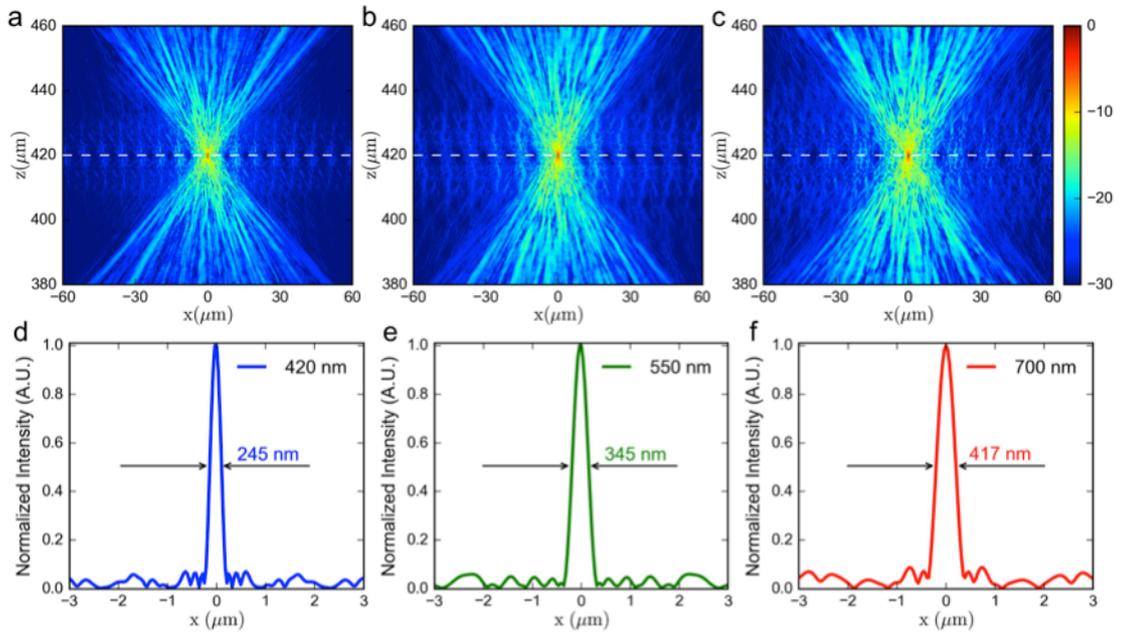

**Figure 4. The focusing pattern of the RGB colors functional metasurface lens.**
(a)-(c) The logarithmic electric field intensity distribution near the focus spot for the wavelength of 420, 550 and 700 nm, respectively. The white dashed line indicates the position of the focal spot, along which the field profile is plotted in (d)-(f) at these wavelengths. The diffraction limited minimum focus spots are obtained here.

**Figure 5 by Yuan *et al.***

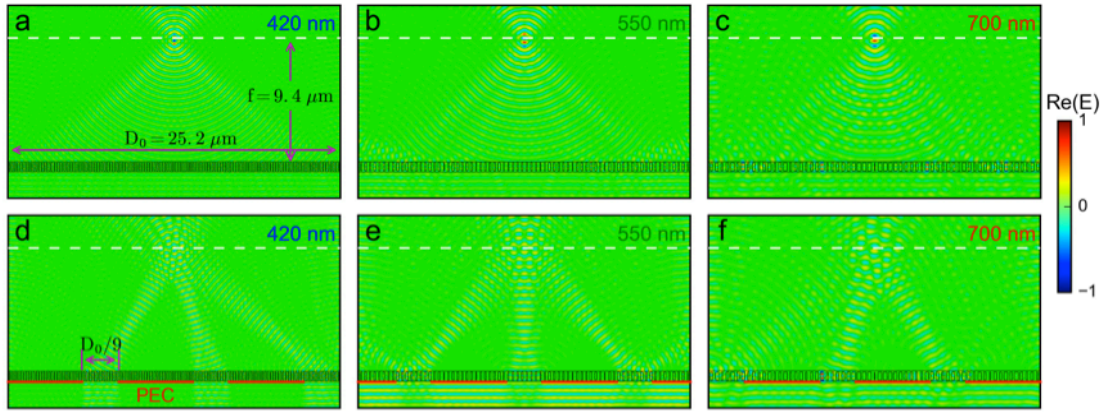

**Figure 5. Numerical simulation of the metasurface lens for RGB colors.** The lens's aperture, focal length and NA are 25.2 μm, 9.4 μm and 0.8, respectively. (a)-(c) The normalized real electric field distribution of the metasurface lens composed of TiO$_2$ rods embedded in equally thick Teflon medium (thickness = 0.8 μm) for normal incident plane wave of wavelength $\lambda_0$ = 420, 550 and 700 nm, respectively. (d)-(f) The electric field distribution with the incident surface of the lens covered by PEC screens left with three open apertures (length = $D_0/9$) at different RGB colors.

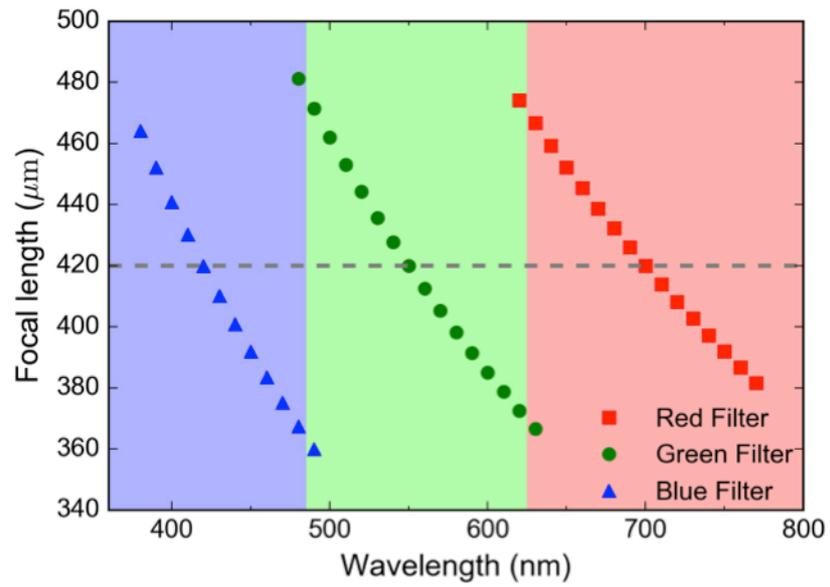

**Figure 6. Chromatic aberration of the RGB colors functional metasurface lens.** It gives the variation of the focal length of the different color sectors designed in the compound lens. The grey dashed line indicates the predesigned focal length for each color sector. The largest focal length variation in the working window is about ± 50 µm.

**Figure 7 by Yuan *et al*.**

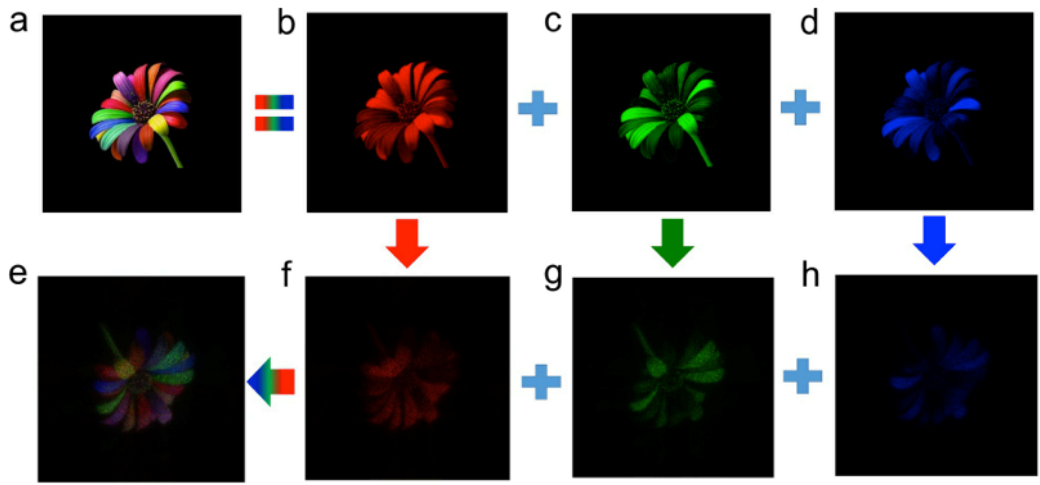

**Figure 7. Imaging application of the proposed RGB colors functional metasurface lens.** The color picture in (a) can be decomposed into red (b), green (c) and blue (d) three single color pictures. After passing the metasurface lens, the imaged pictures for different channels are drawn in (f)-(h) and their composition effect is given in (e). In our calculation, the object and image have an equal distance of two times focal lengths from the metasurface lens.